\pdfoutput=1

\documentclass[11pt]{article}

\usepackage[]{acl}
\usepackage{graphicx}
\usepackage{booktabs}
\usepackage{tabularx}
\usepackage{array}
\usepackage{balance}
\usepackage{tikz}

\usepackage{times}
\usepackage{latexsym}

\usepackage[T1]{fontenc}

\usepackage{tikz}

\newcommand*\circled[1]{\tikz[baseline=(char.base)]{
        \node[shape=circle,draw,inner sep=2pt] (char) {#1};}}
\newcolumntype{L}[1]{>{\raggedright\let\newline\\\arraybackslash\hspace{0pt}}m{#1}}

\usepackage[utf8]{inputenc}

\usepackage{microtype}

\title{Human-in-the-Loop Synthetic Text Data Inspection   \\
with Provenance Tracking}

\usepackage{authblk}
\usepackage{hyperref}
\renewcommand\footnotemark{}

\author[1]{Hong Jin Kang\textsuperscript{*}\thanks{\textsuperscript{*}Authors contributed equally.}}
\author[1]{Fabrice Harel-Canada\textsuperscript{*}}
\author[2]{Muhammad Ali Gulzar}
\author[1]{Violet Peng}
\author[1]{Miryung Kim}
\affil[1]{University of California, Los Angeles}
\affil[2]{Virginia Tech}
\affil[ ]{}
\affil[ ]{\href{mailto:hjkang@cs.ucla.edu}{hjkang@cs.ucla.edu} \hspace{2cm} \href{mailto:fabricehc@cs.ucla.edu}{fabricehc@cs.ucla.edu}}

\newcommand{\tool}{{\sc Inspector}}

\begin{document}
\maketitle
\begin{abstract}

Data augmentation techniques apply transformations to existing texts to generate additional data. 
The transformations may produce low-quality texts, where 
the meaning of the text is changed and the text may even be mangled beyond human comprehension.
Analyzing the synthetically generated texts and their corresponding labels is slow and demanding. 
To winnow out texts with incorrect labels, we develop \tool{}, a human-in-the-loop data inspection technique. 
\tool{} combines the strengths of provenance tracking techniques with assistive labeling.
\tool{} allows users to group related texts by their \textit{transformation provenance}, i.e., the transformations applied to the original text, or \textit{feature provenance}, the linguistic features of the original text.
For assistive labeling, \tool{} computes metrics that approximate data quality,
and allows users to compare the corresponding label of each text against the predictions of a large language model. 
In a user study, \tool{} increases the number of texts with correct labels identified by $3\times$ on a sentiment analysis task and by $4\times$ on a hate speech detection task.
The participants found grouping the synthetically generated texts by their common transformation to be the most useful technique.
Surprisingly, grouping texts by common linguistic features was perceived to be unhelpful. 
Contrary to prior work, our study finds that no single technique obviates the need for human inspection effort. 
This validates the design of \tool{} which combines both analysis of data provenance and assistive labeling to reduce human inspection effort.

\end{abstract}

\begin{figure}
    \centering
    \includegraphics[width=\columnwidth]{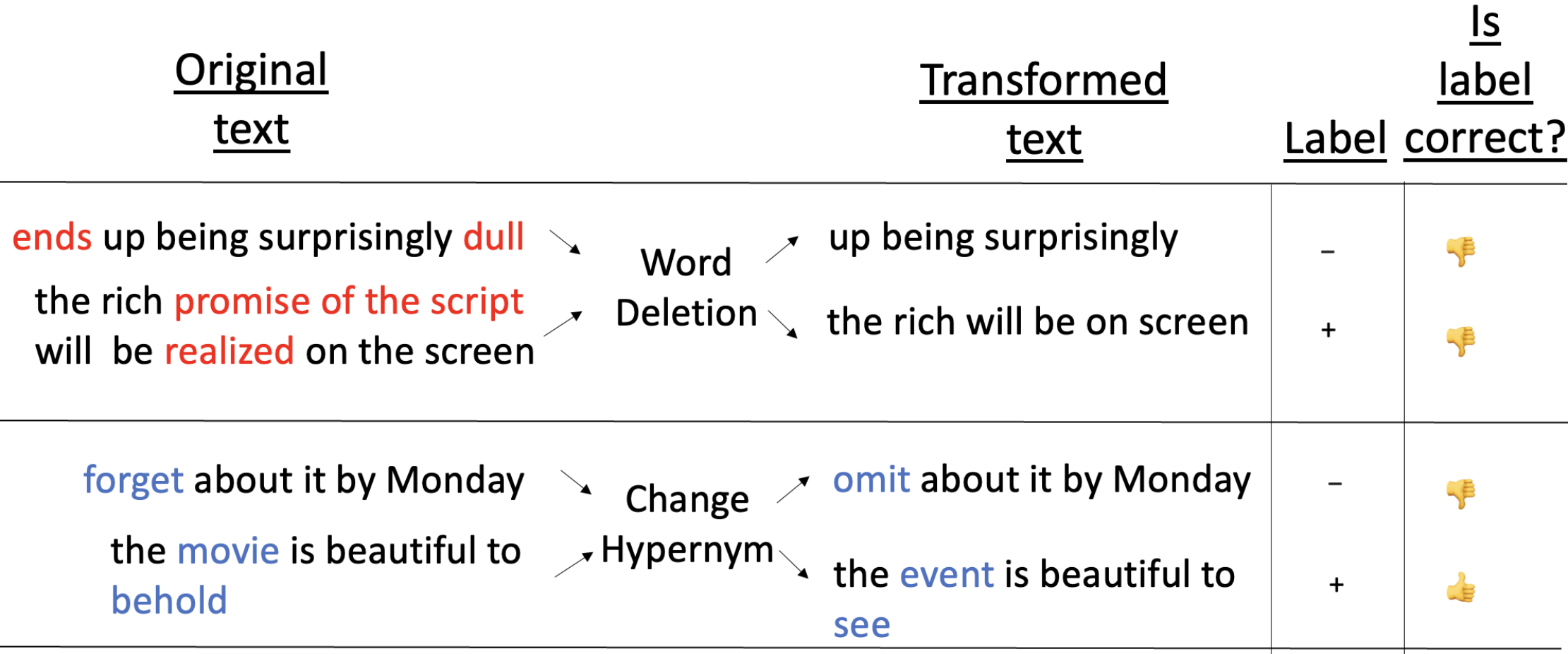}
    \caption{Examples of transformed texts from the SST2 movie review dataset generated during data augmentation. A transformed text can contain garbled text, or have an inappropriate label. As an example, the ``Word Deletion'' transformation can mangle the text ``ends up being surprisingly dull'' into ``up being surprising'', causing its corresponding label ``-'' (indicating a negative sentiment) to no longer be appropriate. Of the four examples of synthetically generated texts, only one (``the event is beautiful to see'') has an appropriate label.}
    \label{fig:introduction}
    \vspace{-3mm}
\end{figure}

\section{Introduction}

\begin{figure*}[!ht]
    \centering
    \includegraphics[width=1.05\textwidth]{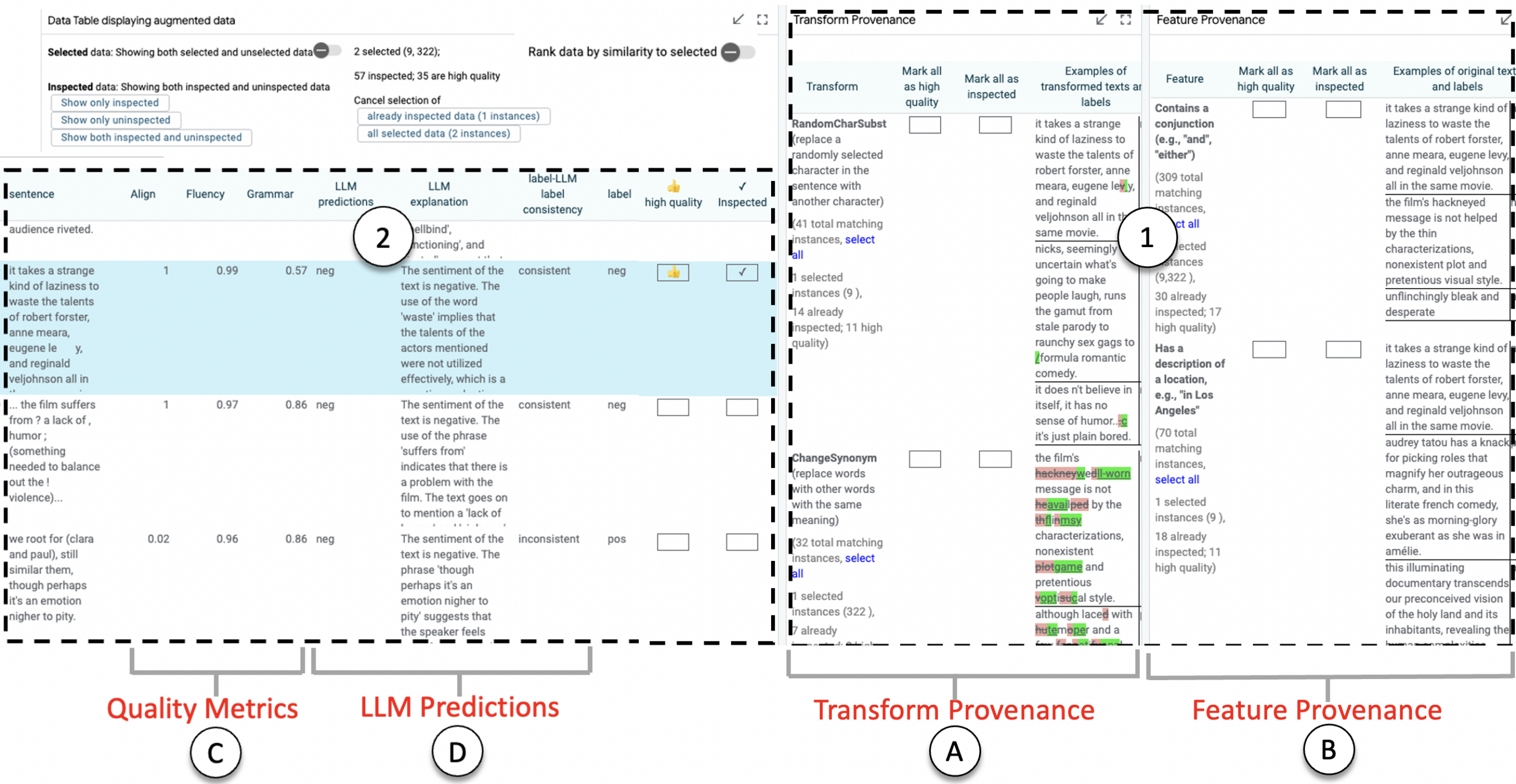}
     \caption{\tool{}: The user alternates between (1) inspecting the provenance of groups of texts and labels following their (A) common transformation, and (B) common linguistic features, and (2) inspecting individual transformed texts with their corresponding labels, with assistive labeling using (C) the quality metrics, alignment, grammaticality, fluency scores, and (D) LLM predictions.}
    \label{fig:workflow}
    \vspace{-3mm}
\end{figure*}

Data augmentation techniques to generate additional training data by transforming existing data can help improve model performance and robustness.
However, low-quality texts with garbled text and inappropriate labels may be generated.
Figure \ref{fig:introduction} shows examples of high- and low-quality instances after a transformation is applied.
Despite users' inclination to filter out texts of low quality with inappropriate labels, effective debugging of the generated content remains challenging due to the opaqueness of these techniques and the sheer volume of data produced. 
Investigating the data instances one by one would be extremely demanding and slow. \looseness=-1

We propose a human-in-the-loop approach, \tool{}, for inspecting generated texts to weed out texts with incorrect labels. 
For reducing human effort, \tool{} applies provenance tracking, inspired by work in the database community~\cite{wang2015big}, and assistive labeling. 
\tool{} supports analysis of the provenance of each text in two ways. 
First, \tool{}  allows users to group the texts by their \textit{transformation provenance}, i.e., the common transformations that have been applied to produce the text.
Second, \tool{} allows texts to be grouped by their \textit{feature provenance}, i.e., common linguistic features, e.g., if the text contains a negation, obtained from the relations represented in Abstract Meaning Representation graphs~\cite{banarescu2012abstract}.
Provenance tracking allows the inspection of groups of texts with the same applied transformations or underlying feature.
For example, the texts in Figure \ref{fig:introduction} transformed by \textit{WordDeletion} share a common transformation and can be grouped together by \tool{}. 

For assistive labeling,  
\tool{} provides two techniques. 
\tool{} computes quality metrics, such as label alignment, grammaticality, and fluency, for each generated text and corresponding label.
Finally, \tool{} provides the predictions from a large language model that users can compare against the corresponding labels of the texts to find discrepancies. 

To evaluate \tool{}, we ran a within-subject user study with 15 participants. 
The participants weeded out generated texts with inappropriate labels on two datasets, a sentiment analysis dataset~\cite{socher2013recursive} and a hate speech detection dataset~\cite{barbieri2020tweeteval}. 
We build a baseline by disabling the 
provenance tracking and assistive labeling features.
Participants using \tool{} identified 3x and 4x more texts with correct labels (Welch’s t-test: p < 0.005).
Using \tool{}, participants were more confident in identifying texts with correct labels, and adopted systematic inspection strategies.
The human-selected texts and labels improve model robustness more than randomly sampled data, demonstrating the value of human inspection.
No single technique of \tool{} was useful to every participant, suggesting that effective inspection of generated texts requires combining complementary techniques.

In summary, 
\tool{} is an approach for inspecting generated texts and corresponding labels using a novel technique for grouping texts by their provenance. 
\tool{} also offers assistive labeling techniques.
Our tool is open source~\cite{dpml_github}.
A within-subject user study shows that using \tool{} enables more effective inspection
and that users found grouping texts by their transformation provenance to be the most useful feature. 
The human-inspected data improves model robustness by up to 32\%.
We find that no single technique, including LLM-based assistive labeling~\cite{gilardi2023chatgpt,wang2021want}, takes away the need for human inspection of generated texts.


\section{Design Goals and System Overview}

\subsection{Design Goals}

Acquiring data is a bottleneck for machine learning~\cite{paleyes2022challenges},
but data labeling is not a simple task~\cite{hansen2013quality,kulesza2014structured,chang2017revolt}.
In particular, ensuring the quality of data is critical ~\cite{liang2022advances,sambasivan2021everyone, paleyes2022challenges}.
We identify challenges for inspecting generated texts:

\textbf{Scale.} 
Data annotation aims to obtain as much data at the lowest possible cost~\cite{wang2022whose}.
Inspection of generated texts shares a similar goal: \tool{} should empower users to identify large subsets of texts with correct labels.

\textbf{Evaluating each instance.} Human inspection is difficult because the data-generating processes are opaque.
Moreover, users may not enjoy trivial labeling tasks~\cite{cakmak2010designing} and 
may clean data without rigor~\cite{krishnan2016towards}.
Hence, \tool{} should discourage ad-hoc inspection and respect human cognition by providing more support for analyzing each text and discovering systematic insights.



\subsection{Overview}

\begin{figure}[!t]
    \centering
    \includegraphics[width=0.5\textwidth]{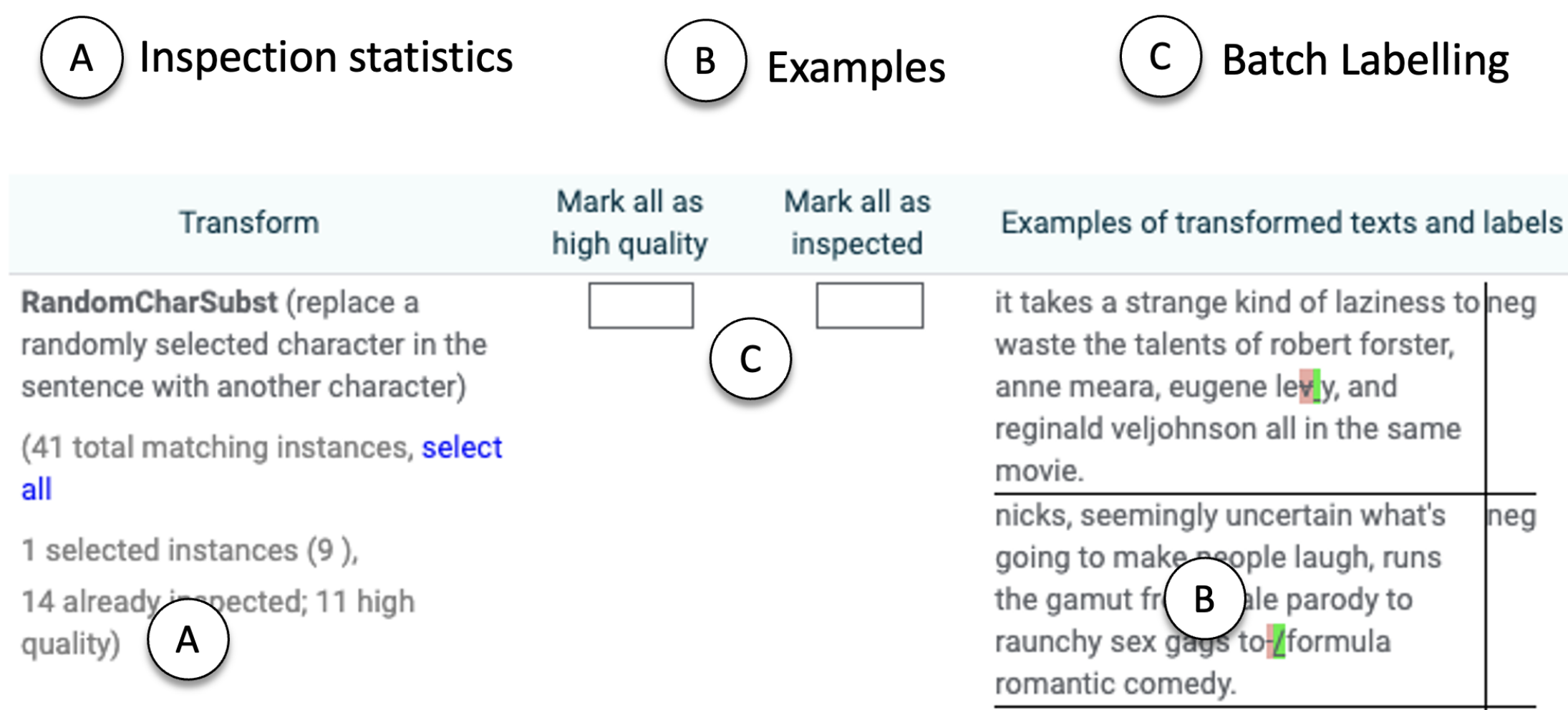}
    \caption{Transform provenance. A user selects texts and inspects the common transforms (e.g., \texttt{RandomCharSubset}) in the transformation provenance pane with their (A) inspection statistics (e.g., the user has inspected 14 texts, with 11 marked as high quality), and (B) view other texts sharing the same transform. A user can then (C) mark all instances sharing the same transform to be correct, obviating the need for inspecting individual texts one by one.}
    \label{fig:transform_provenance}
    \vspace{-3mm}
\end{figure}

\textbf{Workflow.}
\tool{} supports the workflow shown in Figure \ref{fig:workflow}. 
Through \tool{}, users inspect the generated texts and identify texts with correct labels, which are retained in the dataset.
\tool{} enables inspection of the provenance (the two panes (\circled{2}) in Figure \ref{fig:workflow}, Section \ref{sec:provenance_panes}), and assistive labeling (on each generated text and corresponding label (the table (\circled{1}) in Figure \ref{fig:workflow}, Section \ref{sec:data_table}).
For grouping generated texts using provenance tracking, \tool{} offers information about (1) the transformations applied to the selected data in the Transformation Provenance pane, and 
(2) the linguistic features present in the text before transformations were applied in the Feature Provenance pane.

For assistive labeling, \tool{} provides (1) computed quality metrics (i.e. grammaticality, fluency, and label alignment) for each instance, and (2) the predictions of a large language model.

We envision that a user of \tool{} alternates between the inspection of
common transformations and features, forming hypotheses about root causes of quality, 
and the inspection of individual instances with the help of assistive labeling.

\subsection{Usage Scenario.}
Suppose that Alice is a model developer who wishes to expand a training dataset.
Alice turns to data augmentation but is cynical.
She previously observed that these techniques produce large quantities of data with the majority of texts garbled or have unsuitable labels. 
Alice would not trust a model trained with poor quality data (after all, \textit{garbage in, garbage out}). 
As she wishes to retain only texts with correct labels but finds going through all instances onerous, Alice gives \tool{} a try.

\begin{figure}[!t]
    \centering
    \includegraphics[width=\columnwidth]{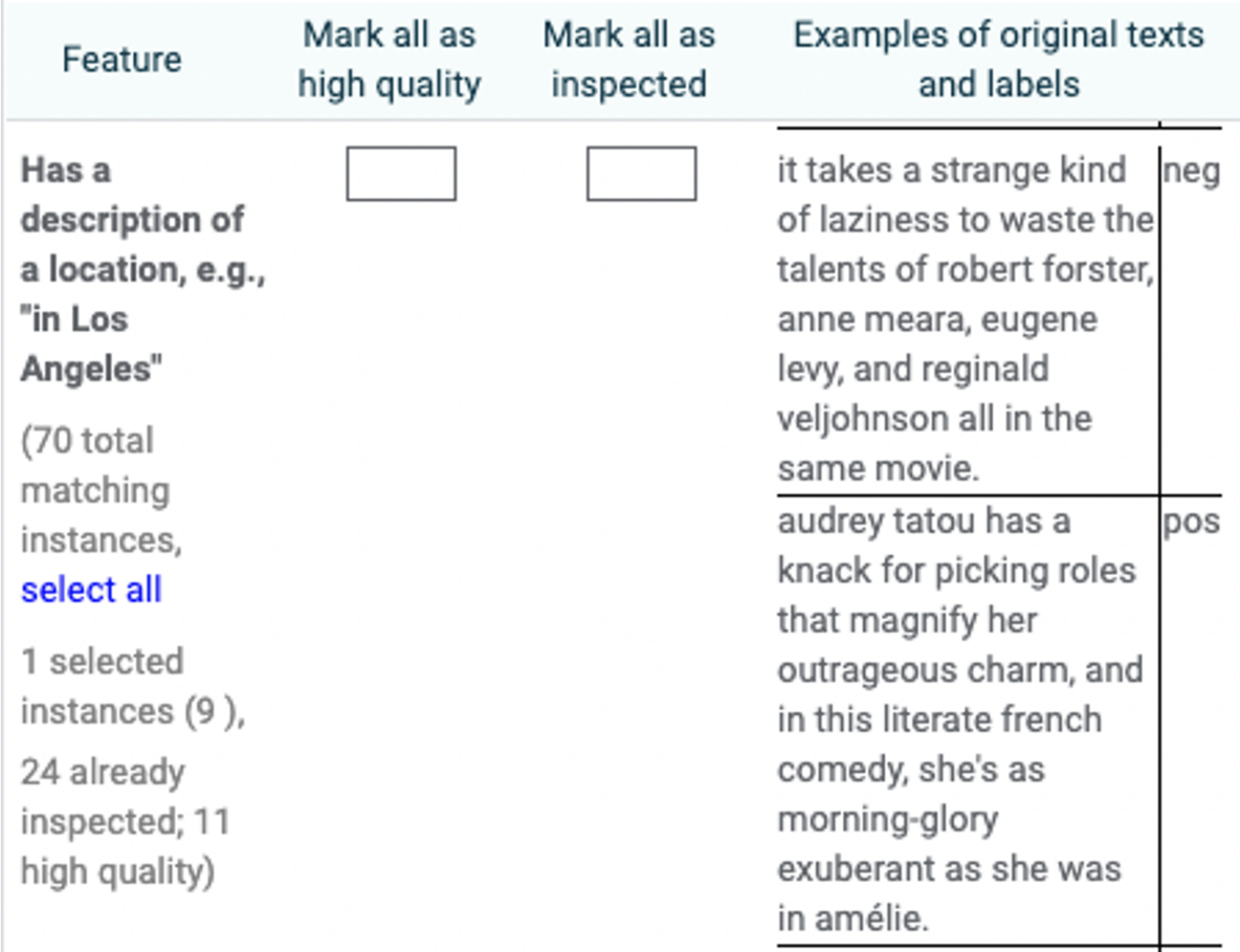}
    \caption{Feature provenance. A user can select texts, and can inspect linguistic features common to the selected texts (e.g., ``Has a description of a location'') in the transformation provenance pane with their inspection statistics (e.g., the user has inspected 24 texts, with 11 marked as high quality).  Then, a user can mark all instances sharing the same feature to be correct. }
    \label{fig:feature_provenance}
    \vspace{-3mm}
\end{figure}

Alice provides the original training dataset to \tool{}.
\tool{} applies the data augmentation techniques to generate data.
Next, Alice inspects the texts (Figure \ref{fig:workflow}). 
First, she inspects several individual instances. 
As she marks their quality, she hypothesizes that the quality of the texts are influenced by a linguistic feature (e.g., in sentiment analysis, deleting a single negation feature in ``I do \textbf{not} like stand-up.'' inverts the sentiment of the text), or a transformation that performs only small-scale modifications.
To assess these hypotheses, she selects several instances of interest. 
Now, the provenance panes display information about the transformations and underlying features of the selected instances (Figure \ref{fig:transform_provenance}).

Alice analyzes the common transformations and underlying linguistic features. 
Each common transformation and feature has examples of data sharing the same provenance pattern, enabling Alice to understand the transformation or feature.
\tool{} tracks and displays inspection statistics such as the total number of inspected records and the proportion of records annotated as having suitable labels.
As Alice continues to mark data, these statistics are updated.
The frequent updates allow Alice to establish trust that the information displayed is meaningful~\cite{lee2004trust,dudley2018review}.
From the small proportion of texts with correct labels among the inspected data (in Figure \ref{fig:feature_provenance}), Alice  
dismisses her hypothesis regarding the linguistic feature. 
However, the same statistic for a common transform in the transformation provenance pane (Figure \ref{fig:transform_provenance}) 
appears to support the hypothesis regarding a transformation that rarely distorts the text.

Alice deepens her investigation. 
She filters the generated texts (Figure \ref{fig:data_pane}) to show only texts sharing the common transformation,
but there is still too much data.
Alice reorders the texts by their grammaticality. 
She finds that even the lowest-scoring instance has a sufficiently high score (e.g., score > 0.8), suggesting that the texts are never too distorted.
Skimming through the texts, she notices that the majority of them have labels that are consistent with the large language model's predictions, suggesting that the transformation usually preserves semantics.
With her hypothesis validated, she marks all instances in the group.
She notices that some instances with incorrect labels have poor fluency scores.
She reorders the texts by their fluency and unmarks the instances with low fluency scores.

Having identified a strategy of finding common transformations among high-quality instances and assessing the grouped instances with the large language model's predictions and quality scores,
Alice continues until she believes she has enough data for training the model.

\section{Implementation}
\subsection{Overview}

\begin{figure}[!htbp]
    \centering
    \includegraphics[width=0.5\textwidth]{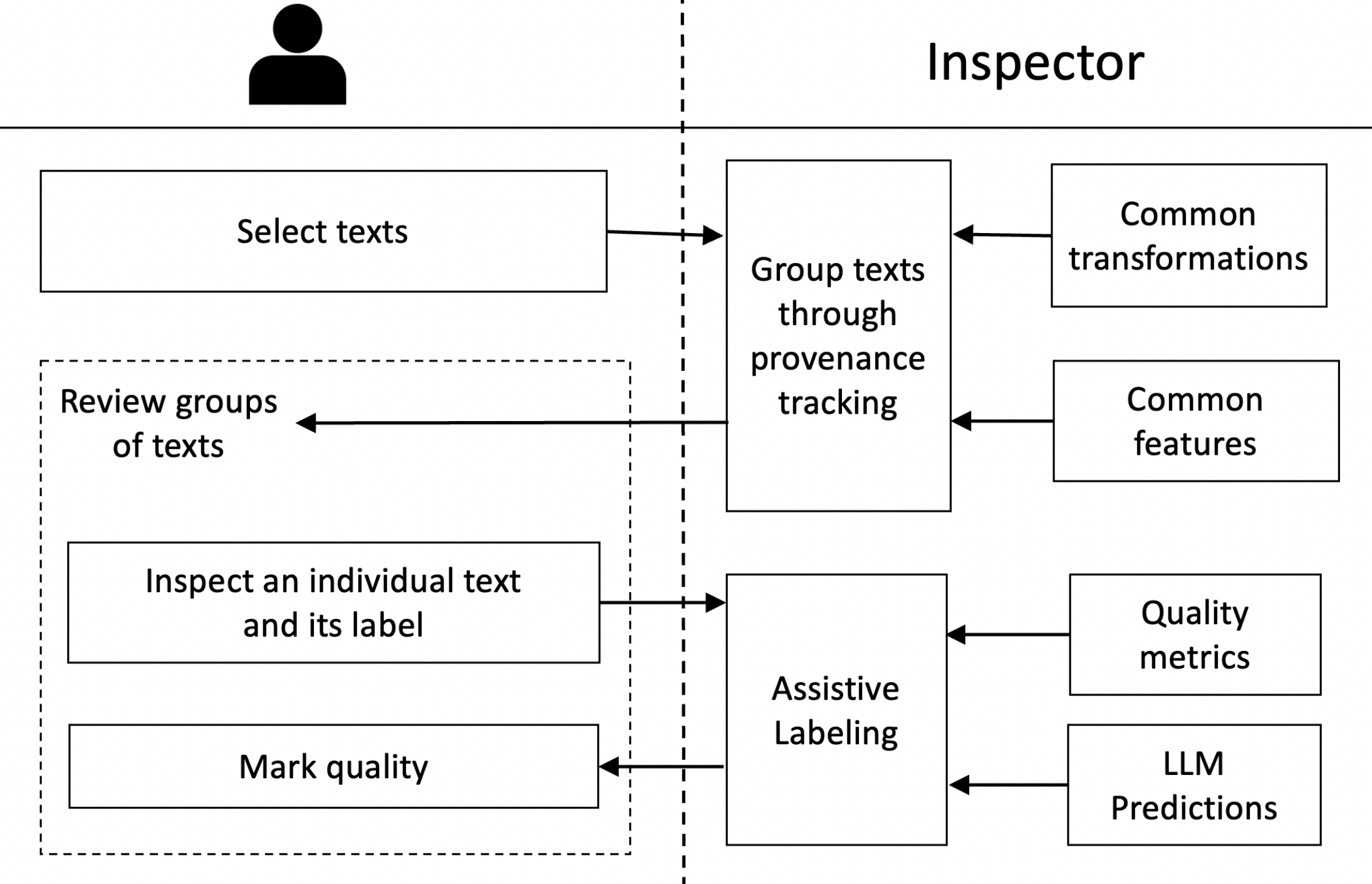}
     \caption{The workflow of a user inspecting data using \tool{}.}
    \label{fig:block}
    \vspace{-2mm}
\end{figure}

Figure \ref{fig:block} shows an overview of how a user interacts with \tool{}.
When a user has selected several texts, \tool{} displays both the synthetically generated texts, as well as their common transformations and features (Section \ref{sec:provenance_panes}).
For each individual text instance, \tool{} computes quality metrics and displays the predictions of a large language model (Section \ref{sec:data_table}).

\subsection{Provenance Tracking}
\label{sec:provenance_panes}

The Transform Provenance and Feature Provenance Panes (Figure \ref{fig:transform_provenance} and Figure \ref{fig:feature_provenance}) surfaces common transformations and linguistic features in the provenance of the user-selected generated texts. 
Providing details about the transforms and features, \tool{} allows users to inspect groups of texts sharing either the same transformations or linguistic features.

The Transformation Provenance pane summarizes recurring patterns in the transformations applied to generate the data.
This allows the investigation of  root causes of poor quality related to  the text transformations, e.g., a random word deletion.
The Feature Provenance pane displays common linguistic features (extracted from Abstract Meaning Representation graphs~\cite{banarescu2012abstract}) in the texts before they were transformed.
This allows the investigation of possible root causes of low quality stemming from features, e.g., negations (``not''), in original texts.


The inspection statistics are presented for each common transform or feature (in the example in Figure \ref{fig:transform_provenance}, the user has inspected 14 texts in total, with 11 of them marked as high quality), enabling the user to make generalizations about the group of data.
Examples of other instances in the group are presented, with the transformed parts of the text highlighted to focus human attention, which may increase labeling efficiency~\cite{choi2019aila}.
Users can filter data to view other instances with a shared transformation or feature provenance.

\textbf{Batch inspection.} Users can mark a batch of instances~\cite{ashktorab2021ai} on a group of data (e.g., all data produced by the same transformation type).
This enables greater inspection efficiency by applying the same decision across the group.

\subsection{Assistive labeling}
\label{sec:data_table}

\begin{figure*}[t]
    \centering
\includegraphics[width=0.9\textwidth]{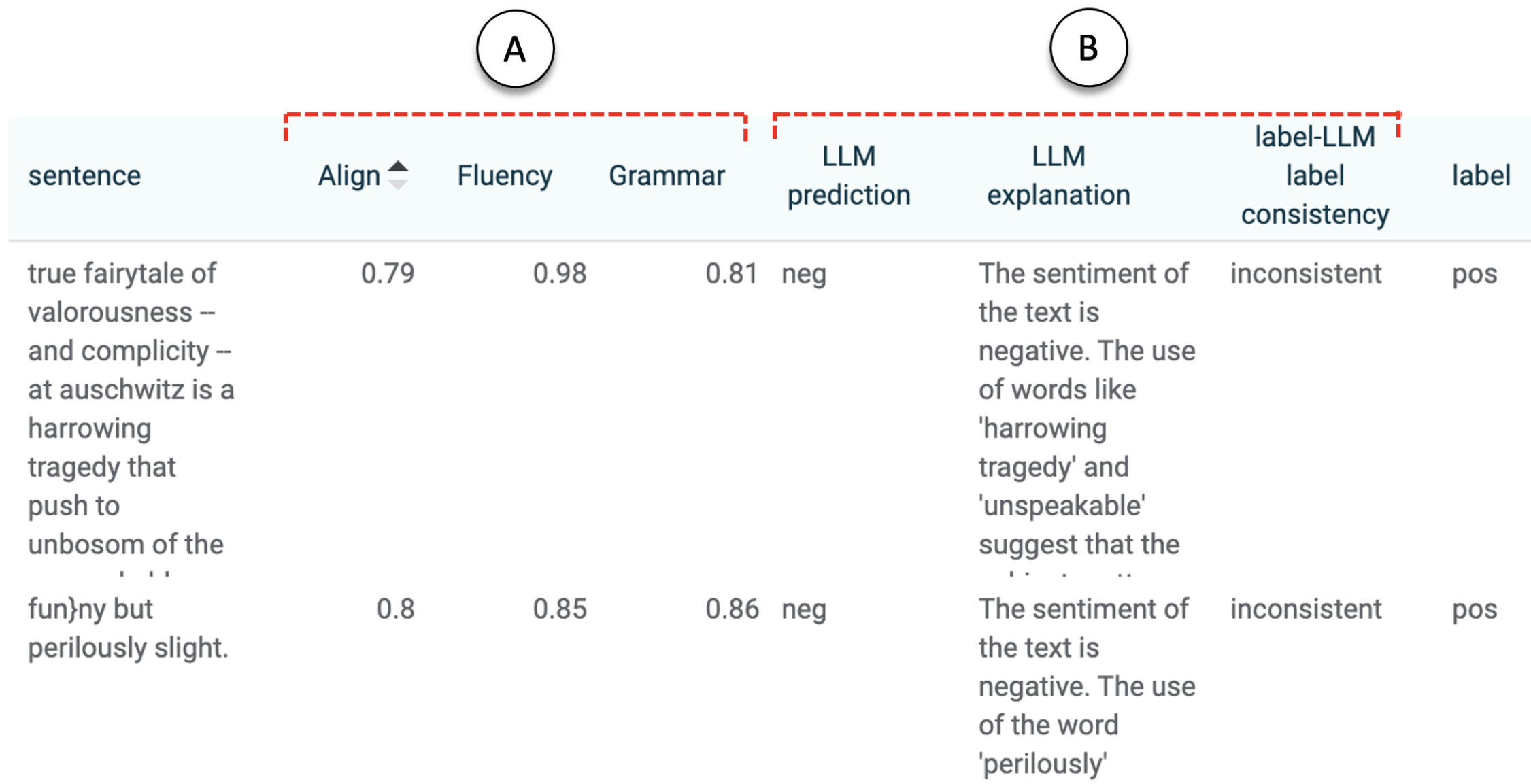}
    \caption{\tool{} enables assistive labeling by providing (A) quality metrics such as label alignment, fluency, and grammaticality. Label alignment is a measure of label quality, while fluency and grammaticality are measures of linguistic quality. It also shows the (B) LLM predictions for a text compared to its corresponding label.}
    \label{fig:data_pane}
    \vspace{-2mm}
\end{figure*}



Figure \ref{fig:data_pane} shows the table containing the transformed texts and their corresponding labels.
\tool{} presents the following quality metrics:

\begin{enumerate}
    \item Fluency, measured through language perplexity using GPT-2~\cite{radford2019language},
    \item Grammaticality, measured as the degree to which the text is free from grammar errors using \texttt{language-tool}~\cite{language_tool},
    \item Label Alignment, measured through predictions of label quality~\cite{northcutt2022confident}.
\end{enumerate}

While Fluency and Grammaticality are measures of linguistic quality, 
Label Alignment measures label quality using CleanLab~\cite{cleanlab}, a method for identifying mislabelled data~\cite{northcutt2022confident}.
We normalize each score such that they range between 0 and 1, with 0 indicating the lowest quality and 1 the highest.
\tool{} allows the texts to be sorted by these metrics.

\tool{} provides information about the outputs of a large language model (\textit{gpt-3.5-turbo}) prompted to predict the transformed texts' labels: 1) its predictions, 2) explanations of each prediction, 3) consistency of its predictions with the instances' labels.
Viewing predictions from the LLM, which may have human-level performance~\cite{gilardi2023chatgpt,wang2021want}, can increase labeling efficiency~\cite{lai2019human,desmond2021increasing} and explanations contextualizing each prediction increases the user's trust in them~\cite{bansal2021does}.
Inconsistencies between labels and LLM predictions may imply altered semantics after a text transformation. 
Together with the quality metrics, the outputs of the LLM provide evidence for users to make decisions, guiding them away from ad-hoc assessments.

             
      

\begin{table}[!t]
\centering
\caption{The text transformations for generating data in the user study. For example, given the text ``ends up being surprisingly dull'' with a ``negative'' label, ``Word Deletion'' produces a new text ``up being surprising'' with the same corresponding label.}
\label{tab:transform_types}
\resizebox{\columnwidth}{!}{%
\begin{tabular}{p{0.28\columnwidth} L{0.72\columnwidth}}
\toprule
\textbf{Category} & \textbf{Transformation} \\ 
\midrule
Swap & ChangeHypernym, ChangeHyponym, ChangeLocation, ChangeName, ChangeNumber, ChangeSynonym, RandomSwap, RandomSwapQwerty \\
\midrule
Punctuation & ContractContractions, ExpandContractions, InsertPunctuationMarks \\
\midrule
Typos & HomoglyphSwap, WordDeletion, RandomCharDel, RandomCharInsert, RandomCharSubst, RandomCharSwap \\
\midrule
Text Insert & RandomInsertion \\
\midrule
Emojis & AddNeutralEmoji, RemoveNeutralEmoji  \\
\bottomrule
\end{tabular}%
}
\end{table}

\section{User Study}

We conducted a within-subject study with 15 participants to assess the effectiveness of  \tool{} for weeding out low-quality texts. 
We developed \tool{} as a web application.
As a baseline, we developed a variant of \tool{} without the effort-reduction techniques.
We investigate the following questions:
\begin{enumerate}
    \item Does \tool{} increase efficiency in identifying texts with correct labels?
    \item How useful is each effort reduction technique offered by \tool{}?
    \item Are models more robust when trained using data identified using \tool{}?
\end{enumerate}

\subsection{Study Design}

\textbf{Tasks.}
The study involves two datasets, the sentiment analysis dataset, SST2~\cite{socher2013recursive}, and a hate speech detection dataset, TweetEval~\cite{barbieri2020tweeteval}, described in Table \ref{tab:datasets}.
Table \ref{tab:transform_types} shows the list of considered text transformations~\cite{morris2020textattack,ribeiro2020beyond,karimi2021aeda,wei2019eda,harel2022sibylvariant}.
For each task completed by the user, we finetune BERT~\cite{devlin2018bert} on the identified high-quality data  for up to 10 epochs.

      

\begin{table}[!t]
\centering
\caption{The two datasets in the user study. We analyze only the synthetic texts generated using the data augmentation techniques. During the course of the user study, 100\% and 99\% of the SST2 and TweetEval datasets were reviewed by at least one participant.}
\label{tab:datasets}
\resizebox{\columnwidth}{!}{%
\begin{tabular}{p{0.2\columnwidth} p{0.5\columnwidth} p{0.1\columnwidth}}
\toprule
\textbf{Dataset} & \textbf{Description} & \textbf{Size} \\
\midrule
SST-2 & Predict the sentiment of a movie review.  & 613 \\
\midrule
TweetEval (Hate) & Predict if a tweet contains offensive discourse. & 763 \\
\bottomrule
\end{tabular}%
}
\vspace{-2mm}
\end{table}

\textbf{Participants.}
We recruited 15 participants by reaching out to students in the Computer Science department.
11 of them are Ph.D. students, 2 are master students, and 2 are undergraduates.
7 participants had less than 1 year of machine learning experience, 2 had about one year, 2 had 2-5 years of experience, and 2 had more than 5 years of experience.
The participants also self-reported their familiarity with inspecting machine learning datasets and understanding mislabelled data on a 7-point Likert scale.
The mean familiarity was 3.4, where 1 is ``Most unfamiliar'' and 7 is ``Most familiar''.
This level of experience is identical to data annotators in industry, who do not have a machine learning background~\cite{wang2022whose}.

\textbf{Study Protocol.}
Our study involves two datasets and two tools.
We design a within-subjects study where each participant investigates both datasets and experiences using both tools.
For each task, the participant used only the assigned tool, either \tool{} or Annotator, the variant of \tool{} without the effort reduction techniques.
The study requires the completion of 2 tasks, with each task requiring 20 minutes at most.
We design our study to be completed in 60 minutes.
Before the users started on a task, we asked for the participants' consent to record their usage of the tools.
Then, they spent 10 minutes working through a tutorial and warm-up questions. 
After completing the tasks, the participants were directed to a post-study questionnaire to share their experiences and feedback about the tools. 
We also solicited responses about the participants' inspection strategy.

\section{Results}
In this section, we report and analyze the results of our user study.
We denote each participant as P\#.

\begin{figure}[!t]
    \centering
    \includegraphics[width=0.5\textwidth]{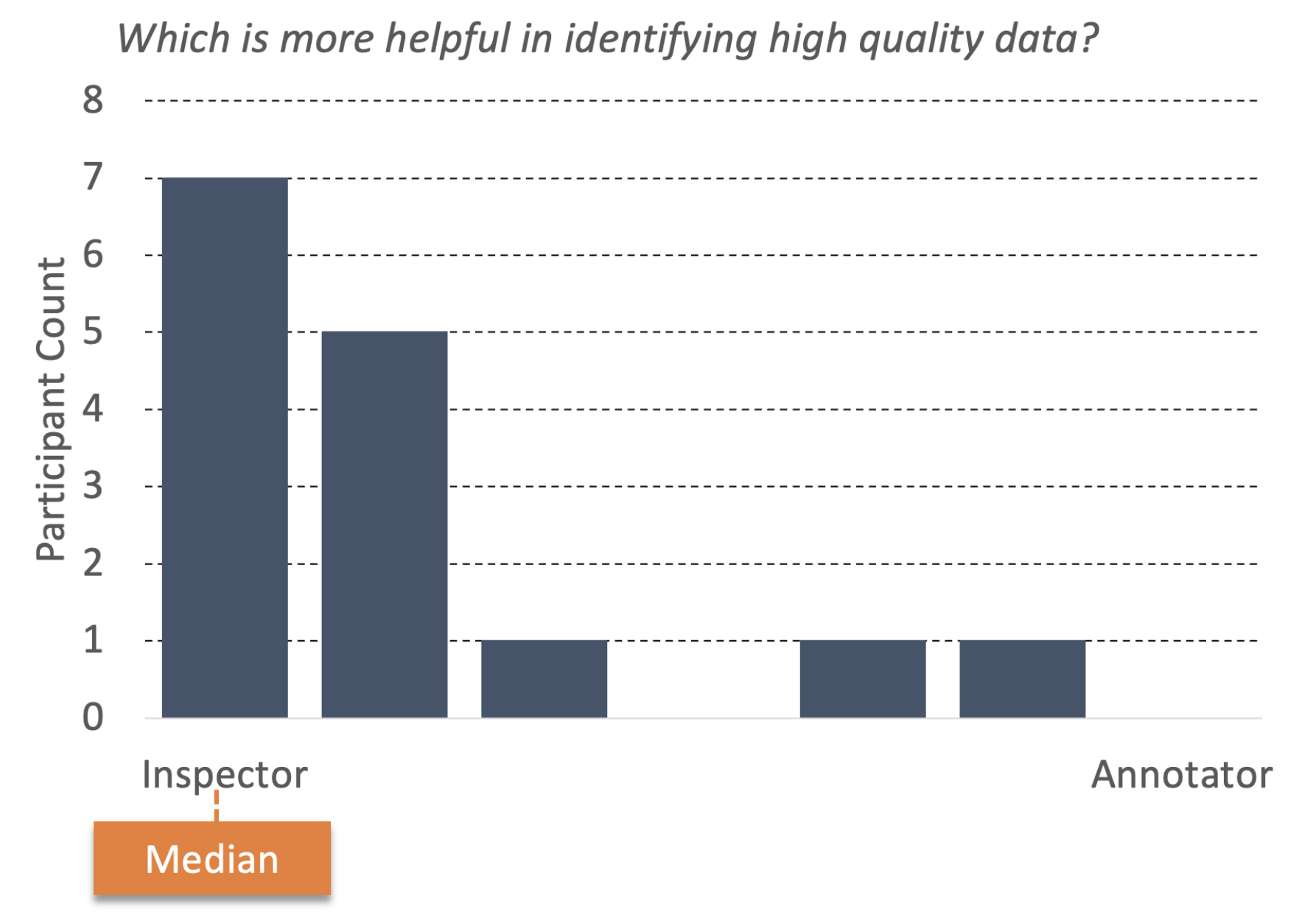}
    \caption{Most participants perceived \tool{} to be more helpful in winnowing out generated texts with incorrect labels (such as the examples in Figure \ref{fig:introduction}).}
    \label{fig:dpml_vs_inspector_likert}
\end{figure}

\begin{figure}[!t]
    \centering
    \includegraphics[width=0.5\textwidth]{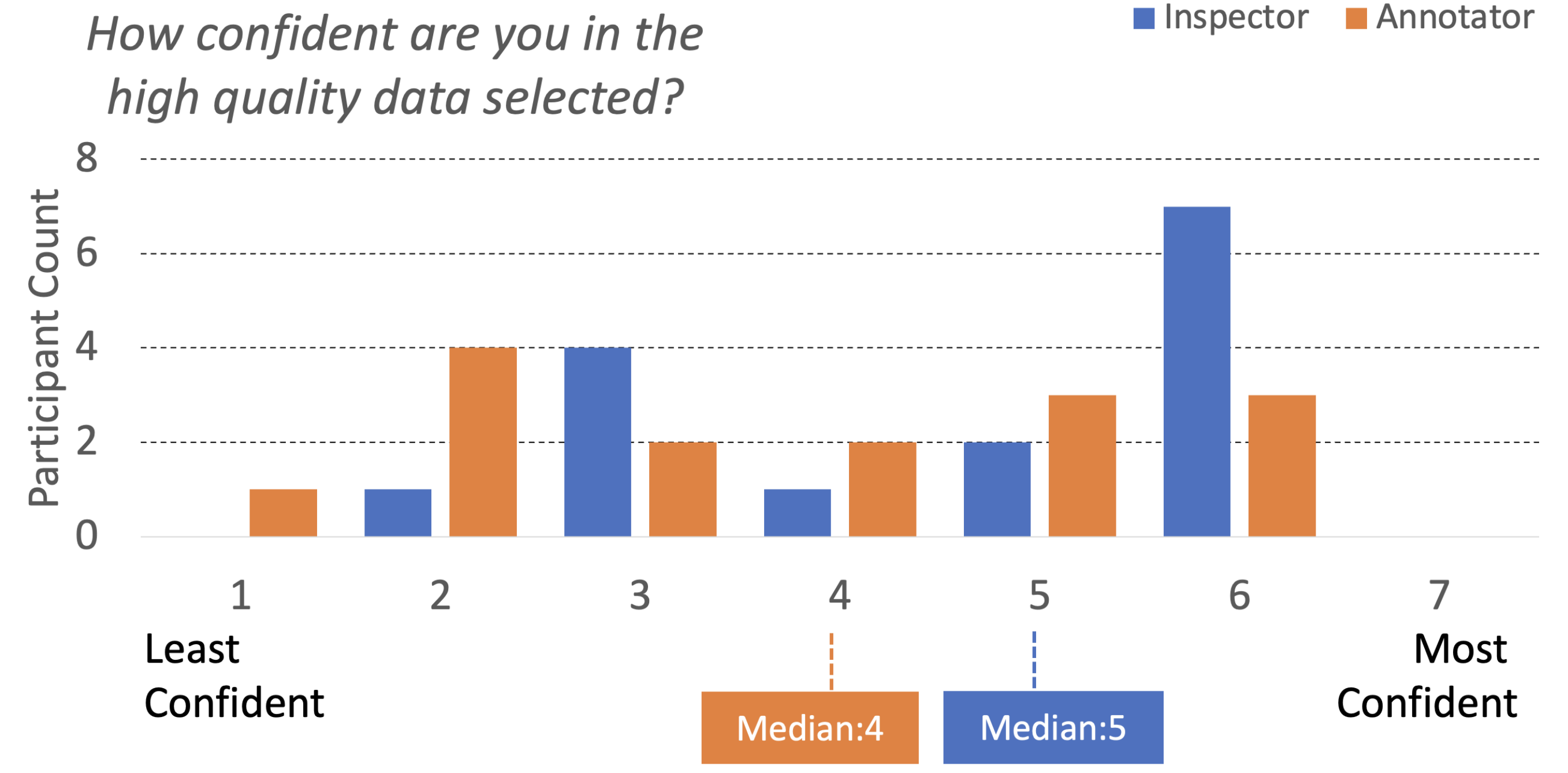}
    \caption{Participants using \tool{} were more confident of their inspections, with a median rating of 5 compared to the baseline of 4.}
    \label{fig:confidence}
\end{figure}


\subsection{Reduction in Human Effort}
To assess inspection efficiency, we counted the number of texts with correct labels identified by the participants. 
Table \ref{tab:cost} shows that using \tool{} leads to 3x and 4x more texts on the sentiment analysis dataset and the hate speech detection dataset. 
Using \tool{}, participants identified an average of 277 and 259 high-quality instances compared to 82 and 63 instances using the baseline on the SST2 and TweetEval's Hate Speech dataset, respectively.
The differences in efficiency is significant (Welch's t-test, p < 0.005).


      

\begin{table}[!t]
\centering
\caption{The average number of texts with correct labels identified. \tool{} improves over a baseline without provenance tracking and assistive labeling by $3\times$ and $4\times$. SA: Sentiment Analysis, HS: Hate Speech Detection}
\label{tab:cost}
\resizebox{\columnwidth}{!}{%
\begin{tabular}{p{0.8\columnwidth} p{0.1\columnwidth} p{0.1\columnwidth}}
\toprule
\textbf{Approach} & \textbf{SA} & \textbf{HS}   \\ 
\midrule
\tool{} w/o provenance tracking and assistive labeling  & 82 & 63 \\
\tool{} & 277 & 259 \\
\bottomrule
\end{tabular}%
}
\vspace{-3mm}
\end{table}

Figure \ref{fig:dpml_vs_inspector_likert} shows that the participants found \tool{} more useful and were more confident of their results.  
13 of the 15 participants indicated that they preferred using \tool{} over the baseline tool. 
Participants had a higher median level of confidence in their inspections using \tool{} (5 vs 4 on a 7-point scale, in Figure \ref{fig:confidence}).

The techniques in \tool{} changed the participants' perception of the task's nature.
P3 indicated that \tool{} ``could give initial results. I'm more acting like a verifier''.
They found \tool{} more usable in identifying patterns.
Comparing \tool{} and the baseline, 13 of the 15 participants found \tool{} more helpful in identifying patterns for inspecting the texts. 
They reported using the techniques of \tool{} in their inspection strategies.
P8 wrote ``My main strategy was to find inconsistent patterns in already labeled data and LLM'' while  P1 mentioned that ``Sentences with a grammar score < 0.92 are almost always low-quality.''.
Conversely, using the baseline, participants found strategically inspecting the data difficult. 
P8 wrote that dissecting the data ``is utterly not possible'' and ``Using \texttt{ctrl+f} was really painful.''

\subsection{User Ratings of Individual Features}


In the post-study questionnaire, participants rated the techniques of \tool{} and described how they inspected the data.
Table \ref{tab:complementary} summarizes the participants' feedback.
Grouping data by transformation provenance was the most appreciated technique, 
followed by the assistive labeling techniques.
On a 7-point Likert scale, participants assigned Transform Provenance the highest average rating of 4.5, followed by the two assistive labeling techniques with average ratings of 4.3. 
P8 indicated that using transform provenance was the ``main strategy I used to identify trends''.
Participant C7 wrote that provenance tracking allows her to ``reason about whether a specific transformation can lead to a reduction of data quality''.


      

\begin{table}[!t]
\centering
\caption{Usefulness of the techniques of \tool{} rated by the participants (out of 7). A higher rating indicates the technique was perceived to be more useful for inspection. Grouping texts by common transformations was perceived to be the most useful technique. }
\label{tab:complementary}
\resizebox{\columnwidth}{!}{%
\begin{tabular}{p{0.7\columnwidth} p{0.2\columnwidth}}
\toprule
\textbf{Technique} &  \textbf{Average Rating}   \\ 
\midrule
Transform Provenance & 4.5 \\
Quality Metrics (grammaticality, \\ fluency, \& alignment) & 4.3 \\
LLM Guidance & 4.3 \\
Feature Provenance & 1.9 \\
\bottomrule
\end{tabular}%
}
\vspace{-3mm}
\end{table}

\textbf{Diverse inspection strategies.} The responses to the post-study questionnaire reflected a wide range of strategies.
No single technique of \tool{} was found to be useful by every participant.
We qualitatively analyze the free responses.
A majority (11) of the participants used the transformation provenance to make decisions.
9 participants described using the quality metrics (grammaticality, fluency, label alignment), and  8 participants mentioned using outputs of the large language model.
All participants found at least one technique of \tool{} to be useful (rating at least 4 out of 7 and using it in their workflow).
This suggests no one technique obviates the need for human inspection.

\textbf{Batch Inspection.}
A majority (11 of the 15) of the participants used the batch inspection feature, marking an average of 5 groups. 
The participants appreciated batch inspection and used it after manual validation of a few representative instances.
P13 wrote ``if most of the labels are not affected by the transformation and are indeed high-quality by manual inspection, I mark all such entries as high-quality.''.
However, not all participants found grouping the texts by their provenance useful or meaningful.
Two participants applied a strategy of trying to go through the data one by one. 
P11 wrote that she did not consider data provenance for grouping the texts ``in order to give an unbiased opinion'' for each text and its label.

\subsection{Robustness}


\begin{table}[!t]
\centering
\caption{Results comparing the robustness of BERT after finetuning on additional data. We measure the attack success rate of DeepWord (the lower the better). SA: Sentiment Analysis, HS: Hate Speech Detection}
\label{tab:robustness}
\resizebox{\columnwidth}{!}{%
\begin{tabular}{p{0.8\columnwidth} p{0.1\columnwidth} p{0.1\columnwidth}}
\toprule
\textbf{Approach} & \textbf{SA} & \textbf{HS}   \\ 
\midrule
Randomly sampled &  0.61 & 0.50 \\
Human-guided & 0.59 & 0.34 \\
\bottomrule
\end{tabular}%
}
\vspace{-2mm}
\end{table}

Next, we assess the improvements in model robustness when training a model with the texts with correct labels collected by the participants. 
As a baseline, we construct a randomly sampled dataset with the average number of data instances selected using \tool{}.
We assess the robustness of the model by measuring the attack success rate of DeepWord~\cite{gao2018black}, a method of generating adversarial attacks.
Using the implementation of DeepWord in TextAttack~\cite{morris2020textattack},
we generated 100 attacks on each model.
More robust models would face fewer successful attacks.

On the SST2 dataset, 4 out of 8 participants marked data that led to more robust models than randomly sampled data.
On the TweetEval dataset, all 7 participants identified data that led to more robust models than randomly sampled data.
The attack success rate of DeepWord on models trained with randomly selected data was 0.61 on the SST2 dataset and 0.5 on the TweetEval dataset.
Using the inspected data, the attack success rate decreases to an average of 0.59 and 0.34 on SST2 and TweetEval, respectively.
On TweetEval, this corresponds to a 32\% improvement.
Overall, the texts identified using \tool{} improved model robustness.
 


\section{Implications}

Our study showed that \tool{} empowered users to be effective and confident in inspecting transformed texts and their corresponding labels.


\textbf{Transformation provenance was the most useful technique.} 
The participants perceived transformation provenance to be the most useful technique provided by \tool{}. 
Even when participants did not use it to perform batch inspection, they found the additional information useful.
P2 considered this information to inspect data, writing that it ``showed whether a malformed sentence was in the original text or mangled by a transformation.''.

\textbf{Assistive labeling helped users build trust.} 
The participants were aware of the risk of including incorrect labels when labeling a batch of texts.
Thus, the users had to trust the guidance and automation provided by \tool{}.
Users were able to build trust using the inspection statistics of the groups of generated texts.
The assistive labeling techniques were also helpful for building confidence. 
P8 wrote ``When the fluency and grammaticality scores are both high, I am more confident to label the data as high quality.''


\textbf{The linguistic features were perceived to be ineffective.} 
Surprisingly, users found grouping the texts by their common linguistic features to be ineffective. 
The participants did not always understand their relevance to the task. 
P14 wrote ``the features are very low level and it is not clear how they are related to the labeling quality''.

\textbf{Limited impact on inspection accuracy.}
We observed that users tended to only mark texts they were confident about.
Comparing the participants' inspections to the ground-truth labels annotated by one of the authors of this paper, we find that the use of \tool{}  only had a limited impact on the accuracy of the user's inspections. 
On SST2, \tool{} slightly decreased labeling accuracy from 88\% to 85.3\%.
On TweetEval, \tool{} increased accuracy from 89.1\% to 90.0\%. 
These differences are not statistically significant (p-value $> 0.05$).
This suggests that users of \tool{} inspected more texts with less effort and increased confidence while maintaining the same level of accuracy.

\section{Related Work}

\subsection{Debugging Machine Learning}

\textbf{Testing models.}
DynaBench~\cite{kiela2021dynabench} evaluates models with challenging human-provided data. 
AdaTest~\cite{ribeiro2022adaptive} generates more test cases similar to user-provided tests. 
These studies focus on models but do not support debugging data, which is the focus of \tool{}.

\textbf{Grouping data for data inspection.}
SliceFinder \cite{chung2019automated} and What-If~\cite{wexler2019if} 
are tools for understanding the subsets of data with poor model performance.
Zeno~\cite{cabrera2023zeno} groups data on properties extracted by user-written Python programs. 
Tempura~\cite{wu2020tempura} supports  data inspection by 
grouping data using structural templates 
abstracted from concrete data instances.
Unlike these tools, \tool{} groups data by 
their provenance.

\subsection{Cleaning Data}
For cleaning tabular data, Data Civilizer~\cite{rezig2019data} is a data pipeline-debugger for identifying low-quality data (e.g., malformed values) that cause incorrect data analysis outputs through inserting breakpoints and the visualization of the intermediate records. 
Unlike Data Civilizer, \tool{} is not a breakpoint debugging tool. Instead, it helps users filter out data with unsuitable labels.
Wrangler~\cite{kandel2011wrangler} cleans dirty data by inferring data types (e.g., integers) and semantic roles (e.g., zip code) for identifying anomalous data. 
Potter’s Wheel~\cite{raman2001potter} cleans dirty data but detecting them may require users to implement an API to define constraints (e.g., date formats). 
On the other hand, \tool{} assists human users in interactively identifying texts with incorrect labels without requiring programming literacy.

Ruler~\cite{choi2021tagruler} and TagRuler~\cite{evensen2020data} learn rules for annotating unlabelled data based on the text. 
\tool{} guides users to weed out texts with incorrect labels. 
While feature provenance in \tool{} is similar to the token-based rules in Ruler or TagRuler, it uses linguistic features rather than tokens.

\section{Conclusion}
We present \tool{} for winnowing synthetic texts with incorrect labels generated during data augmentation.
In a within-subject user study, 
participants using \tool{} were $3\times$ and $4\times$ more effective.
Users found that grouping data by their shared common transformations to be the most useful technique.
Assistive labeling allowed them to build trust in the tool.
Surprisingly, users perceived the linguistic features to be ineffective. 
\tool{} is the first interactive human-in-the-loop approach for examining text augmentation data for classification tasks by combining provenance inspection and assistive labeling techniques.

We publicly release \tool{}. \tool{} is available at \href{https://github.com/UCLA-SEAL/ProvenanceInspector}{https://github.com/UCLA-SEAL/ProvenanceInspector}.

\section{Limitations}

\tool{} guides human users using several techniques, including the computation of quality metrics and the use of  Abstract Meaning Representation for analyzing feature provenance.
Due to these dependencies, \tool{} is only applicable to texts in languages that are supported by the tools for computing the metrics (CleanLab, LanguageTool) and converting the text to Abstract Meaning Representation.

\tool{} was evaluated through a user study. 
Although our study was performed with student participants, the participants in our study share a similar expertise level with full-time data annotators in industry --- i.e. data annotators in industry are not full-time data scientists or engineers and they generally do not have any machine learning background. 
Consequently, having student participants is unlikely to impact generalizability.

While we evaluated \tool{} on only two datasets, 
\tool{} does not use compute-intensive techniques and would work on large datasets. 
As the design of \tool{} is not specific to the tasks in the user study, we believe that our findings would generalize to other NLP tasks.

Labeling hate speech is known to be inherently ambiguous and influenced by the annotator’s beliefs even when labeling guidelines are provided. 
\tool{} does not solve the issue of annotation bias.

Our work investigated only human effort in inspecting texts generated as part of data augmentation. 
Our insights may, therefore, be specific to data augmentation.
We hope to extend this analysis to other data inspection tasks.

\section{Ethics Statement}

Our work aims to allow human users better insights into generated data. 
These data may contain toxic, offensive content, and \tool{} may expose 
these content to its users. 
Alone, \tool{} does not generate biased or offensive text, however, it postprocesses the output of data augmentation techniques which may produce harmful texts.

We obtained an exemption from the UCLA IRB to run the user studies. 

\section*{Acknowledgements}
This work is supported by the National Science Foundation under grant numbers 2106838, 1764077, 1956322, and 2106404. It is also supported by funding from Amazon and Samsung, the Meta Sponsor Research Award, the Okawa Foundation Research Grant, and the Amazon Alexa AI Research Award. We thank Akshay Singhal for customizing the initial \tool{} interface and Jiayue (Coraline) Sun for developing the early pipelines for provenance tracking and model training. We want to thank the anonymous reviewers for their constructive feedback that helped improve the work.


\bibliography{anthology,custom}




\end{document}